# Infrared Spectra and Chemical Abundance of Methyl Propionate in Icy Astrochemical Conditions

B. Sivaraman[1,*], N. Radhika[2], A. Das[3], G. Gopakumar[2,4], L. Majumdar[3], S. K. Chakrabarti[5,3], K. P. Subramanian[1], B. N. Raja Sekhar[6], M. Hada[2,4]

[1] *Space and Atmospheric Sciences Division, Physical Research Laboratory, Ahmedabad, India.*

[2] *Department of Chemistry, Tokyo Metropolitan University, 1-1 Minami-Osawa, Hachioji, Tokyo, 192-0397, Japan.*

[3] *Indian Centre for Space Physics, 43 Chalantika, Garia Station Road, Kolkata 700084, India.*

[4] *JST, CREST, 4-1-8 Honcho, Kawaguchi, Saitama, 332-0012, Japan*

[5] *S. N. Bose National Centre for Basic Sciences, Salt Lake, Kolkata 700098, India*

[6] *B-4, Indus-1, BARC Spectroscopy Lab at Indus-1, Atomic and Molecular Physics Division, BARC, Mumbai & RRCAT, Indore 452013, India.*

**ABSTRACT**

We carried out an experiment in order to obtain the InfraRed (IR) spectra of methyl propionate ($CH_3CH_2COOCH_3$) in astrochemical conditions and present the IR spectra for future identification of this molecule in the InterStellar Medium (ISM). The experimental IR spectrum is compared with the theoretical spectrum and an attempt was made to assign the observed peak positions to their corresponding molecular vibrations in condensed phase. Moreover, our calculations suggest that methyl propionate must be synthesized efficiently within the complex chemical network of the ISM and therefore be present in cold dust grains, awaiting identification.

***Keywords: Astrochemisry, methods : laboratory : solid state, ISM : abundances, ISM : molecules, infrared : general.***

*Corresponding author: bhala@prl.res.in

## INTRODUCTION

The identification of methyl acetate ($CH_3COOCH_3$) in the ISM (Tercero et al. 2013) implied the presence of propionic acid ($CH_3CH_2COOH$) in similar regions where methyl acetate was detected. Methyl acetate dissociation carried out experimentally simulating the interstellar cold conditions also pointed out the formation of alcohol, methyl alcohol ($CH_3OH$), along with other products to form from acetate (Sivaraman et al. 2014). In Interstellar chemistry, methyl alcohol is among the most dominant molecule influencing various chemical reactions that takes place on cold dust grains. In methyl acetate dominant regions, there are strong chances of propionic acid to be present (Karton & Talbi 2014). The addition of methyl alcohol may lead to the next generation of molecule that can form from the reaction of propionic acid and methyl alcohol; which on the surface of a cold dust grain can lead to the formation of methyl propionate ($CH_3CH_2COOCH_3$) molecule (Equation 1). However, there are no experimental IR spectra available in the literature for methyl propionate at conditions commensurate to cold ISM environment. Therefore, here we present the first experimental IR spectra, along with the theoretical spectra, of methyl propionate that will help in the identification of methyl propionate in various regions of the ISM where methyl propionate must be component as shown by our gas-grain chemical model.

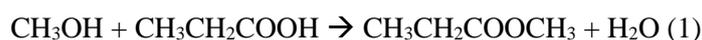

$$CH_3OH + CH_3CH_2COOH \rightarrow CH_3CH_2COOCH_3 + H_2O \quad (1)$$

## METHODS AND THEORETICAL DETAILS

### Experimental method

Experiments were carried out in the low temperature astrochemistry apparatus housed in the Physical Research Laboratory (PRL); a detailed description of the experimental apparatus can be found elsewhere (Sivaraman, et al. 2014). Upon achieving UltraHigh Vacuum (UHV) of the order of $10^{-9}$ mbar and low temperature (85 K) at the zinc selenide (ZnSe) substrate, methyl propionate molecules in the gas phase, extracted from the liquid sample that was freeze pump thawed three times to eliminate any impurity, was let into the chamber to form icy layers of methyl propionate at 85 K.

Sample deposition was monitored by Fourier Transform InfraRed (FTIR) spectrometer operated in the mid-infrared wavelength, 4000 cm$^{-1}$ – 400 cm$^{-1}$. Absorbance scale in the spectra recorded while depositing the sample was used as an indicator to cease the deposition; in this case absorbance was kept less than 1.

**Computational details**

The electronic structure calculations were performed using GAUSSIAN 09 software package (Gaussian 2009). The optimization of molecular geometry and the IR frequency analysis were first carried out at *ab-initio* MP2 (second order Möller-Plesset theory) level (Møller & Plesset 1934). To determine the accuracy of the ab-initio calculation, we also performed the vibrational analysis using DFT B3LYP (Becke 1988; Lee, Yang, & Parr 1988; Stephens et al. 1994) and PBEPBE (Perdew, Burke, & Ernzerhof 1996) functional. All the calculations were performed using the Dunning's correlation consistent basis set cc-pVTZ (Woon & Dunning 1993).

For all the 36 vibrational modes, harmonic and anharmonic frequencies with their corresponding infrared intensities have been computed at DFT as well as MP2 levels of theory using the algorithms implemented in GAUSSIAN 09.

**RESULTS AND DISCUSSION**

To the best of our knowledge, there was no X-ray crystallographic data available for methyl propionate molecule. Hence, we modelled the molecule in gas phase (Figure 1) and obtained the structural parameters. The bond length, bond angle and dihedral angle are tabulated in Table I. Comparing different methods, we find bond lengths to be shorter for MP2 in comparison with the DFT calculations. The bond angles and dihedral angles are found to be similar in both the methods. As methyl propionate can be formed from methyl alcohol and propionic acid as given in Equation 1, we compared the trend in structural parameters among these molecules. The C–H, C–C and C–O bonds were found to be differing by 0.13 Å, 0.01 Å and 0.04 Å respectively (Strieter et al. 1962; Venkateswarlu & Gordy 1955). The C=O bond length varies only by 0.02 Å. Further, the bond angles are changing within 1-5°, thereby reconfirming our structure to be close to experimental values.

**Table I**. Structural parameters of methyl propionate at the DFT-PBEPBE, DFT-B3LYP and MP2 level of theory.

| Method | PBEPBE | | B3LYP | | MP2 | |
|---|---|---|---|---|---|---|
| | Bond length (Å) | | | | | |
| | C1-C4 | 1.5242 | C1-C4 | 1.5228 | C1-C4 | 1.5180 |
| | C1-H2 | 1.0973 | C1-H2 | 1.0891 | C1-H2 | 1.0871 |
| | C1-H3 | 1.0973 | C1-H3 | 1.0891 | C1-H3 | 1.0871 |
| | C1-H14 | 1.0979 | C1-H14 | 1.0900 | C1-H14 | 1.0877 |
| | C4-C7 | 1.5146 | C4-C7 | 1.5117 | C4-C7 | 1.5055 |
| | C4-H5 | 1.1014 | C4-H5 | 1.0926 | C4-H5 | 1.0903 |
| | C4-H6 | 1.1014 | C4-H6 | 1.0927 | C4-H6 | 1.0903 |
| | C7-O12 | 1.2152 | C7-O12 | 1.2049 | C7-O12 | 1.2106 |
| | C7-O13 | 1.3624 | C7-O13 | 1.3503 | C7-O13 | 1.3501 |
| | C8-O13 | 1.4424 | C8-O13 | 1.4362 | C8-O13 | 1.4335 |
| | C8-H9 | 1.0978 | C8-H9 | 1.0887 | C8-H9 | 1.0867 |
| | C8-H10 | 1.0978 | C8-H10 | 1.0887 | C8-H10 | 1.0836 |
| | C8-H11 | 1.0944 | C8-H11 | 1.0857 | C8-H11 | 1.0867 |
| | Bond angle (°) | | | | | |
| | C1-C4-C7 | 113.0 | C1-C4-C7 | 113.2 | C1-C4-C7 | 112.2 |
| | H2-C1-C4 | 111.0 | H2-C1-C4 | 111.1 | H2-C1-C4 | 110.8 |
| | H3-C1-C4 | 111.0 | H3-C1-C4 | 111.1 | H3-C1-C4 | 110.8 |
| | H14-C1-C4 | 110.6 | H14-C1-C4 | 110.4 | H14-C1-C4 | 110.5 |
| | C4-C7-O12 | 125.9 | C4-C7-O12 | 125.8 | C4-C7-O12 | 125.8 |
| | C4-C7-O13 | 110.6 | C4-C7-O13 | 110.9 | C4-C7-O13 | 110.8 |
| | H5-C4-C1 | 111.6 | H5-C4-C1 | 111.4 | H5-C4-C1 | 111.7 |
| | H6-C4-C1 | 111.6 | H6-C4-C1 | 111.4 | H6-C4-C1 | 111.7 |
| | C7-O13-C8 | 114.6 | C7-O13-C8 | 115.8 | C7-O13-C8 | 113.9 |
| | H9-C8-O13 | 110.6 | H9-C8-O13 | 110.6 | H9-C8-O13 | 110.5 |
| | H10-C8-O13 | 105.6 | H10-C8-O13 | 105.7 | H10-C8-O13 | 105.6 |
| | H11-C8-O13 | 110.6 | H11-C8-O13 | 110.6 | H11-C8-O13 | 110.5 |
| | Dihedral angle (°) | | | | | |
| | C1-C4-C7-O13 | 179.9 | C1-C4-C7-O13 | 179.9 | C1-C4-C7-O13 | 180.0 |
| | C8-O13-C7-C4 | 180.0 | C8-O13-C7-C4 | 180.0 | C8-O13-C7-C4 | 180.0 |
| | H2-C1-C4-C7 | 59.7 | H2-C1-C4-C7 | 59.8 | H2-C1-C4-C7 | 59.6 |
| | H3-C1-C4-C7 | -59.6 | H3-C1-C4-C7 | -59.7 | H3-C1-C4-C7 | -59.5 |
| | H14-C1-C4-C7 | -179.9 | H14-C1-C4-C7 | -179.9 | H14-C1-C4-C7 | -180.0 |
| | H9-C8-O13-C7 | -60.3 | H9-C8-O13-C7 | -60.4 | H9-C8-O13-C7 | -60.3 |
| | H10-C8-O13-C7 | 180.0 | H10-C8-O13-C7 | 180.0 | H10-C8-O13-C7 | 180.0 |
| | H11-C8-O13-C7 | 60.3 | H11-C8-O13-C7 | 60.3 | H11-C8-O13-C7 | 60.3 |

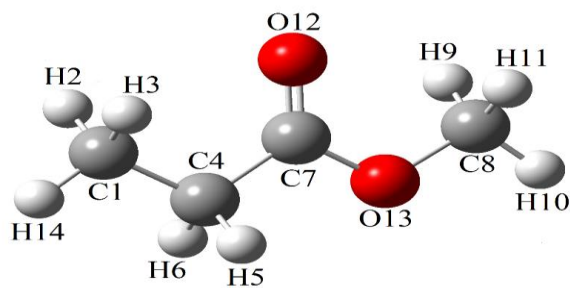

Figure I. Structure of methyl propionate at equilibrium geometry at MP2 level.

In Table II, we present the harmonic and anharmonic IR frequencies obtained at DFT and MP2 levels of theory. In the mid and higher IR frequency regions, we find that the MP2 gives larger frequency in comparison with DFT and therefore, the spectra obtained from MP2 level of theory is shown in Figure II(a).

**Table II.** Calculated harmonic ($\omega$) and anharmonic ($\nu$) frequencies of methyl propionate (in cm$^{-1}$) with corresponding A values (cm molecule$^{-1}$) at MP2 level.

| Method | DFT-PBEPBE | | DFT-B3LYP | | MP2 | | A value |
|---|---|---|---|---|---|---|---|
| Band Assignment | Frequency ($\omega$) | Frequency ($\nu$) | Frequency ($\omega$) | Frequency ($\nu$) | Frequency ($\omega$) | Frequency ($\nu$) | |
| cCH3-rocking(?) | 40 | 98 | 46 | 58 | 39 | 50 | 6.69E-21 |
| cCH3-rocking(?) | 120 | 48 | 130 | 51 | 148 | 142 | 4.58E-19 |
| oCH3-rocking(?) | 146 | 228 | 154 | 160 | 163 | 172 | 3.56E-19 |
| COC bend(?) | 200 | 285 | 207 | 212 | 211 | 215 | 3.67E-19 |
| cCH3-rocking(?) | 200 | 221 | 210 | 280 | 226 | 220 | 3.41E-19 |
| COC bend(?) | 319 | 330 | 330 | 334 | 338 | 362 | 3.10E-18 |
| CCO bend(?) | 435 | 433 | 445 | 442 | 452 | 446 | 1.06E-19 |
| C=O wag | 563 | 560 | 584 | 581 | 585 | 581 | 3.50E-19 |
| OCO bend | 635 | 624 | 656 | 665 | 664 | 666 | 5.61E-19 |
| cCH3-rocking | 785 | 779 | 817 | 809 | 816 | 811 | 1.34E-18 |
| CCO Stretch | 832 | 819 | 859 | 847 | 868 | 853 | 2.36E-18 |
| O-CH3 stretch | 961 | 935 | 979 | 955 | 1008 | 985 | 4.98E-19 |
| CCO Stretch | 1001 | 985 | 1031 | 1014 | 1048 | 1028 | 2.27E-18 |
| CCO bend | 1066 | 1040 | 1106 | 1082 | 1116 | 1092 | 3.50E-20 |
| cCH3-rocking | 1075 | 1048 | 1113 | 1088 | 1127 | 1098 | 1.85E-18 |
| oCH3-rocking | 1130 | 1109 | 1176 | 1150 | 1192 | 1167 | 2.12E-19 |
| skeletal def. | 1153 | 1117 | 1201 | 1168 | 1214 | 1183 | 7.97E-18 |
| skeletal def. | 1166 | 1134 | 1219 | 1179 | 1239 | 1202 | 5.28E-17 |
| cCH3-twist | 1238 | 1199 | 1286 | 1250 | 1293 | 1263 | 7.80E-21 |
| CCO bend | 1327 | 1286 | 1384 | 1345 | 1395 | 1357 | 1.49E-17 |
| cCH3-wag | 1366 | 1332 | 1423 | 1411 | 1424 | 1389 | 4.77E-19 |
| CH2-bend | 1408 | 1361 | 1466 | 1437 | 1474 | 1433 | 1.90E-18 |
| oCH3-wag | 1412 | 1379 | 1474 | 1449 | 1484 | 1449 | 2.55E-18 |
| oCH3-rocking | 1428 | 1390 | 1486 | 1445 | 1506 | 1465 | 1.35E-18 |
| cCH3-rocking | 1440 | 1416 | 1495 | 1485 | 1511 | 1487 | 1.18E-18 |
| oCH3-scissoring | 1446 | 1415 | 1501 | 1480 | 1518 | 1488 | 2.46E-18 |
| cCH3-scissoring | 1450 | 1408 | 1504 | 1483 | 1521 | 1493 | 1.30E-18 |
| C=O stretch | 1745 | 1713 | 1798 | 1766 | 1801 | 1771 | 2.89E-17 |
| sym-cCH3-stretch | 2975 | 2837 | 3038 | 2942 | 3092 | 3026 | 2.75E-18 |
| symoCH3+symcCH2 | 2979 | 2887 | 3045 | 3015 | 3096 | 2987 | 3.95E-18 |
| symoCH3+symcCH2 | 2986 | 2858 | 3049 | 3002 | 3097 | 3012 | 2.39E-18 |
| Asym-cCH2-stretch | 3002 | 2843 | 3062 | 2910 | 3140 | 3000 | 3.45E-19 |
| asymcCH3-stretch | 3049 | 2899 | 3107 | 2980 | 3182 | 3046 | 2.60E-18 |
| asym-cCH2-stretch | 3056 | 2896 | 3114 | 2976 | 3186 | 3049 | 2.28E-18 |
| asym-oCH2-stretch | 3059 | 2900 | 3117 | 2969 | 3188 | 3048 | 2.86E-18 |
| asym-oCH3-stretch | 3087 | 2954 | 3150 | 3002 | 3221 | 3083 | 1.86E-18 |

Spectra recorded after methyl propionate deposition at 85 K (Figure II(b)) showed several peaks originating from the vibrations of the methyl propionate molecule as a result of interaction with IR photons. Many peaks were observed in the 4000 – 400 cm$^{-1}$ region (Table III) and are found to be in good agreement with the bands that are reported in the 654 cm$^{-1}$ – 1360 cm$^{-1}$ region (Moravie and Corset. 1974). Based on the values calculated the peak positions observed 654 cm$^{-1}$, 807.7 cm$^{-1}$, 839 cm$^{-1}$ and 965.2 cm$^{-1}$ are assigned to OCO bend, cCH3 rocking, CCO stretching and O-CH$_3$ stretching vibrations, respectively. Another set of intense bands with peak positions at 1020.6 cm$^{-1}$, 1089 cm$^{-1}$, 1159 cm$^{-1}$ and 1180.5 cm$^{-1}$ are assigned to CCO stretching, c-CH$_3$ rocking vibrations and the last two towards skeletal deformation, respectively. Next set of vibrational bands with peak positions at 1207 cm$^{-1}$, 1328 cm$^{-1}$, 1362.1 cm$^{-1}$, 1378.7 cm$^{-1}$, 1417.8 cm$^{-1}$, 1439.4 cm$^{-1}$ and 1461 cm$^{-1}$ are assigned to cCH$_3$ twisting, CCO bend or skeletal deformation, cCH$_3$ wagging, CH$_2$ bending, oCH$_3$wagging, OCH$_3$ rocking and cCH$_3$ scissoring.

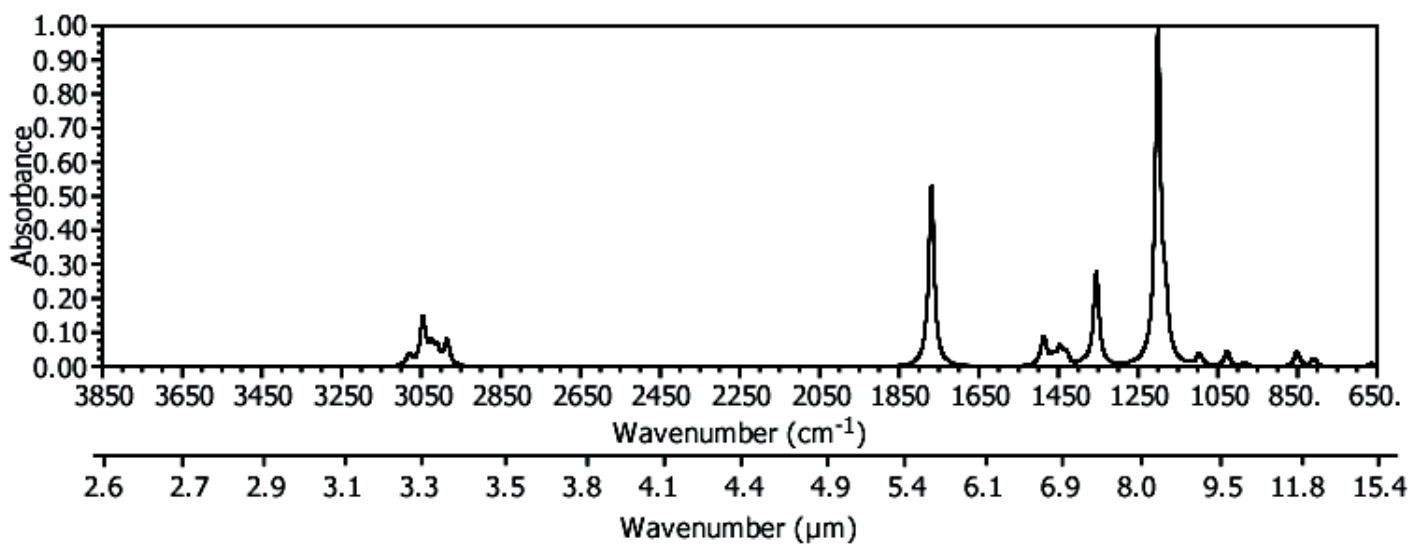

Figure **II. (a)** Calculated infrared spectrum of methyl propionate in gas phase at MP2 level of theory. The IR spectrum has been displayed as a Lorentzian function with a half-width at a half-maximum (hwhm) value of 8cm$^{-1}$ using the Gabedit software (Allouche 2011).

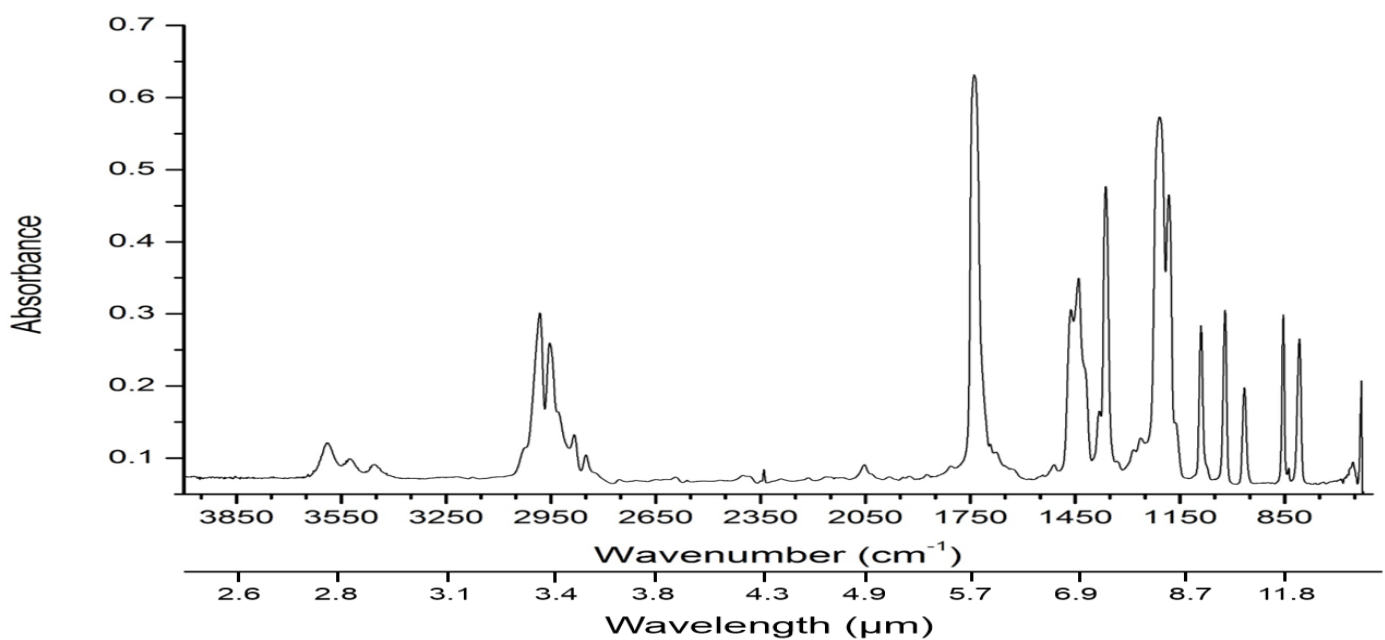

Figure **II. (b)** Infrared spectrum of methyl propionate recorded after deposition at 85 K on to a ZnSe substrate.

Other set of bands with peak positions at 1739.2 cm$^{-1}$, 2965.6 cm$^{-1}$, 2926.9 cm$^{-1}$/ 2951.7 cm$^{-1}$, 2982 cm$^{-1}$ and 3027.7 cm$^{-1}$ are assigned to CO stretching, symmetric cCH$_3$ stretching, combination bands (originating from symmetric oCH$_3$ stretching + symmetric cCH$_2$ stretching), combination band and asymmetric oCH$_3$ stretching vibrations, respectively. However, several other peaks with peak positions at 854.2 cm$^{-1}$, 1262.7 cm$^{-1}$, 1284.7 cm$^{-1}$, 1507.2 cm$^{-1}$, 1804.4 cm$^{-1}$, 2848 cm$^{-1}$, 2883.7 cm$^{-1}$, 3453.7 cm$^{-1}$, 3524.5 cm$^{-1}$ and 3590 cm$^{-1}$ are left unassigned. These bands may correspond to either overtone/combination bands or may even originate from the interaction between two MP molecules (dimers) (Hesse and Suhm. 2009). The contribution from dimer molecules cannot be neglected as it is well known (Sivaraman et al. 2012; Sivaraman et al. 2013) in such low temperature molecular ices where the deposited parent molecule containing H, O and C atoms are known to have different sets of dimers (conformers).

**Table III**. Peak positions observed in the mid-IR spectrum of condensed methyl propionate at 85 K, under astrochemical conditions. Values in microns are given in bracket.

| Wavenumber, cm$^{-1}$ | | |
|---|---|---|
| 654 (15.3) | 1262.7 (7.9) | 1804.4 (5.5) |
| 807.7 (12.4) | 1284.7 (7.8) | 2051.2 (4.9) |
| 839 (11.9) | 1328 (7.5) | 2848 (3.5) |
| 854.2 (11.7) | 1362.1 (7.3) | 2883.7 (3.5) |
| 965.2 (10.4) | 1378.7 (7.3) | 2926.9 (3.4) |
| 1020.6 (9.8) | 1417.8 (7.1) | 2951.7 (3.4) |
| 1089 (9.2) | 1439.4 (6.9) | 2982 (3.4) |
| 1159 (8.6) | 1461 (6.8) | 3027.7 (3.3) |
| 1180.5 (8.5) | 1507.2 (6.6) | 3453.7 (2.9) |
| 1207 (8.3) | 1739.2 (5.7) | 3524.5 (2.8) |
| | | 3590 (2.8) |

Though there are several peaks that correspond to the methyl propionate molecule, the very intense peaks were observed at 1739.2 cm$^{-1}$ and 1159 cm$^{-1}$. The 1739.2 cm$^{-1}$ corresponds to the CO stretch which might not be an unique feature to identify methyl propionate as CO stretch may arise from other molecules containing COO group, such as methyl acetate, and the other very intense peak at 1159 cm$^{-1}$ (8.62 μm) coincides the intense and broad silicate bands. Identification even using the rest of the bands that may fall under the silicate 8 – 12 μm absorption may be difficult. We have to depend on the other peaks to identify this molecule in the cold regions of the interstellar region, in such a case, based on the IR spectra obtained experimentally a set of bands centring at 654 cm$^{-1}$ (15.2 μm), 807.7 cm$^{-1}$ (12.4 μm), 1362 cm$^{-1}$ (7.3 μm), 1439 cm$^{-1}$ (6.94 μm) and 2982 cm$^{-1}$ (3.35 μm) can be used.

**Chemical Modelling for the formation of Methyl Propionate**

In order to study formation of methyl propionate in ISM, we develop a gas-grain chemical model. Our gas phase chemical network contains the network of Woodall et al. (Woodall et al. 2007) and the network used in Das et al. (Das et al. 2013), Majumdar et al. (Majumdar et al. 2013) and Majumdar et al. (Majumdar, Das, & Chakrabarti 2014). We also include a few new gas phase reactions for the formation and destruction of methyl propionate and its related species. Our surface network mainly adopted from Das et al. (Das, et al. 2013) and references therein. Our present gas phase chemical network consists of 6160 reactions among 607 gas phase species and present surface chemical network consists of 285 reactions. Gas phase species are allowed to accrete on the grain surface. In order to mimic dense cloud condition, we consider, $n_H = 10^4$ cm$^{-3}$, T = 10 K and $A_V$ = 10.

**Table IV**. Reactions for the formation and destruction of Methyl Propionate.

| Reaction step | Type of reaction | Rate coefficient |
|---|---|---|
| 1. $CH_3OH + CH_3CH_2COOH \rightarrow CH_3CH_2COOCH_3 + H_2O$ | Neutral-neutral | 1.16 x 10$^{-8}$ cm$^3$ s$^{-1}$ |
| 2. $C_2H_4^+ + CO \rightarrow C_2H_4CO^+$ | Ion-molecular | 3.47 x 10$^{-10}$ cm$^3$ s$^{-1}$ |
| 3. $C_2H_4CO^+ + H_2O \rightarrow C_2H_4CO^+H_2O$ | Ion-molecular | 1.105 x 10$^{-8}$ cm$^3$ s$^{-1}$ |
| 4. $C_2H_4CO^+H_2O + H_2O \rightarrow CH_3CH_2COOH + H_2O$ | Ion-molecular | 1.07 x 10$^{-8}$ cm$^3$ s$^{-1}$ |
| 5. $C_2H_4CO^+ + e^- \rightarrow C_2H_4 + CO$ | Dissociative Recombination | 1.1 x 10$^{-6}$ cm$^3$ s$^{-1}$ |
| 6. $C_2H_4CO^+H_2O + e^- \rightarrow C_2H_4CO^+ + H_2O$ | Dissociative Recombination | 1.1 x 10$^{-6}$ cm$^3$ s$^{-1}$ |
| 7. $CH_3CH_2COOH^+ + e^- \rightarrow C_2H_5 + CO + OH$ | Dissociative | 8.21 x 10$^{-7}$ cm$^3$ s$^{-1}$ |

|   | | Recombination |   |
|---|---|---|---|
| 8. $CH_3CH_2COOH + PHOTON \rightarrow CH_3CH_2COOH^+ + e^-$ | | Photo-dissociation | $2.0 \times 10^{-10}$ s$^{-1}$ |
| 9. $CH_3CH_2COOH + CRPHOT \rightarrow C_2H_5 + CO + OH$ | | CRP | $4.05 \times 10^{-15}$ s$^{-1}$ |

Gas phase methyl propionate mainly forms through reactions between propionic acid ($CH_3CH_2COOH$) and methyl alcohol ($CH_3OH$) (Reaction no. 1). But the main challenge in chemical modelling is the unavailability of reaction pathways of its related species as well as required rate coefficients. For this, we prepare a complete reaction network for the formation and destruction of Methyl propionate and its related species as shown in Table IV.

Reaction number 1 is the main gas phase production route of methyl propionate. But rate coefficient of this reaction is yet to be known. To this effect, we carried out quantum chemical calculation to find out rate coefficient of this reaction. We then find out the properties of this reaction. For this, we optimize the geometry of the reactants and products separately by using B3LYP functional with 6-311G++(d,p) basis set. It is noticed that the difference between total enthalpy of products and total enthalpy of reactants is –ve (i.e., the reaction is spontaneous in forward direction), so reaction number 1 is exothermic in nature. To compute reaction rate of reaction 1, we use dispersive form of capture rate theory (Clary et al. 1994):

$$k_{NN} = 8.56 \, C_6^{1/3}/\mu^{1/2} \, (K_B T)^{1/6} \quad cm^3 \, molecule^{-1} \, s^{-1} \quad (R1),$$

where, $\mu$ is collisional reduced mass, $K_B$ is the Boltzmann constant and $C_6 = 3/2(E_aE_b/E_a+E_b) \, \alpha_a\alpha_b$, where, $E_a$ and $E_b$ are ionization potentials and $\alpha_a$ and $\alpha_b$ are the polarizabilities of $CH_3OH$ and $CH_3CH_2COOH$ respectively. We use, $E_a = 10.85$ eV, $E_b = 10.525$ eV and $\alpha_a = 3.29 \times 10^{-24}$ cm$^3$ and $\alpha_b = 6.9 \times 10^{-24}$ cm$^3$ (Lide 2001). If all the parameters in equation (R1) are used in atomic units then the factor $0.613 \times 10^{-8}$ converts it into units of cm$^3$ molecule$^{-1}$ s$^{-1}$ (Clary, et al. 1994). Using all the parameters, we have the rate coefficient $1.16 \times 10^{-8}$ cm$^3$ s$^{-1}$ for the Reaction no. 1.

Propionic acid is the basic ingredient for the formation of methyl propionate in the gas phase. Following Blagojevic et al. (Blagojevic, Petrie, & Bohme 2003), we use some Ion-molecular reactions (Reaction nos. 2-4) for the formation of propionic acid. Rate coefficients of these reactions were till date unknown. According to the discussion section of Majumdar et al. (Majumdar, et al. 2014) and references therein, Langevin collision rate ($k_L$) formula could be used to compute rate coefficients of ion neutral reactions, provided that neutrals are non-polar:

$$k_L = 2\pi e \sqrt{\alpha_d/\mu} \; cm^3 \, s^{-1}$$

But in case of reaction number 2, 3 and 4 (Table IV), both the neutrals ($H_2O$ and CO) are polar in nature, so the situation would be different. Su & Chesnavich, (Su & Chesnavich 1982) and Woon & Herbst (Woon & Herbst 2009) used parameterized trajectory theory to compute the rate coefficients of this type of reactions. According to Woon & Herbst, (Woon & Herbst 2009), the Su-Chesnavich formula can be written in two different ways and both of which uses a parameter

$$x = \mu_D / \sqrt{2\alpha K_B T}$$

where $K_B$ is the Boltzmann constant, $\mu_D$ is the dipole moment of the polar neutral species in Debye and T is the temperature in Kelvin. Depending on the values of 'x', rate coefficients could be parametrized by following two equations;

$$k_{IN} = (0.4767x + 0.6200) \, k_L \text{ for } x \geq 2,$$

$$k_{IN} = [(x + 0.5090)2/10.526 + 0.9754] \, k_L \text{ for } x < 2.$$

Note that for x = 0, above equations reduce to the Langevin expression. Following a similar approach, here also, we calculate the rate coefficients of these Ion-polar reactions.

Methyl propionate is a stable molecule, but its related species could be destroyed by various mechanism, which could affect its production. Cationic species are mainly destroyed by the Dissociative recombination (DR) reactions shown in Table IV. DR pathways for $C_2H_4CO^+$ (Reaction no. 5) and $C_2H_4CO^+H_2O$ (Reaction no. 6) are chosen by following the DR pathways of $CH_2CO^+$ (Woodall, et al. 2007) and calculated by using following rate formula :

$$k_{DR} = \alpha \, (T/300)^\beta \exp(-\gamma/T) \; cm^3 \, s^{-1},$$

where, $\alpha$, $\beta$, $\gamma$ are the three constants. By following the rate constants used for $CH_2CO^+$, here also, we use $\alpha = 2 \times 10^{-7}$, $\beta = -0.5$, $\gamma = 0$ for Reaction nos. 5 and 6. DR pathways for $CH_3CH_2COOH^+$ (Reaction no. 7) has been assigned by following DR pathways available for HCOOH (Woodall, et al. 2007) and use $\alpha = 1.5 \times 10^{-7}$, $\beta = -0.5$, $\gamma = 0$.

Propionic acid could be ionized by the interstellar photon. Following the reaction of HCOOH (Woodall, et al. 2007), here we consider Reaction no. 8. Rate coefficient of this reaction is calculated by,

$k_{PH} = \alpha \exp(-\gamma A_V)$ s$^{-1}$,

where, $A_V$ is the visual extinction parameter. For the dense cloud condition, we use $A_V = 10$. For Reaction no. 8, we use $\alpha = 2.6 \times 10^{-10}$ and $\gamma = 2.6$ by following the pathways of HCOOH in Woodall et al. (2007).

Cosmic ray induced photo-dissociation (CRP) effect also dissociate Propionic acid. Reaction number 9 is considered here by following the photo-dissociation pathways of HCOOH (Woodall, et al. 2007). Rate coefficient of reaction number 9 is calculated by,

$k_{CR} = \alpha (T/300)^\beta \gamma/(1-\omega)$,

where, $\omega$ is the dust grain albedo in the far ultraviolet (we use $\omega = 0.6$ and T = 10 K). For this reaction we use, $\alpha = 1.3 \times 10^{-17}$, $\gamma = 124.5$ by following the pathways for HCOOH (Woodall, et al. 2007).

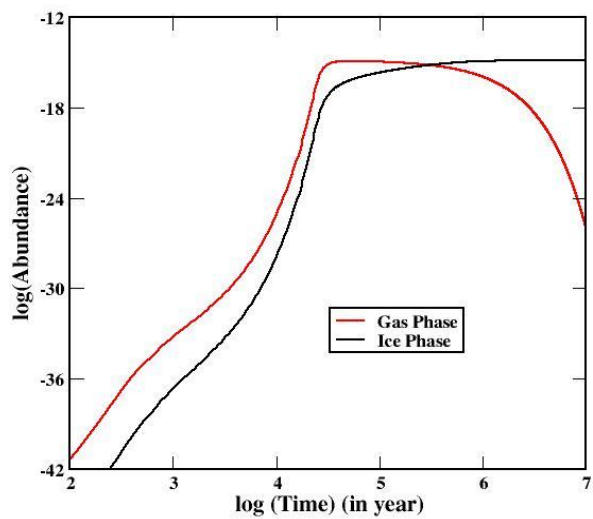

Figure **III**. Time evolution of gas and ice phase methyl propionate in interstellar conditions.

In Figure. III, we have shown the chemical evolution of methyl propionate in gas phase as well as in the ice phase. Here, we are not considering any formation or destruction pathways for the ice phase methyl propionate. But since, we are considering depletion of gas phase species, Interstellar grains could be populated by methyl propionate as well. Since no destruction pathways for the ice phase methyl propionate are considered, it is achieving a steady state during the late stages. Abundance of gas phase methyl propionate decreases, due to the destruction of its related species and depletion to the interstellar grain. From Figure. III, we observe a peak abundance of methyl propionate with respect to total hydrogen nuclei to be $1.2 \times 10^{-15}$ and $1.3 \times 10^{-15}$ respectively for the gas and ice phase.

## CONCLUSION AND IMPLICATIONS

Experimental IR spectra obtained showed several bands that corresponds to condensed methyl propionate at 85 K. Though there are several bands that shows the characteristic vibrations from methyl propionate molecule, it is only those bands falling on either side of the strong and broad silicate absorption that are presented in this paper can be used to detect its presence in the ISM. However, future experimental work to be carried out on methyl propionate ices in binary mixtures, such as methyl propionate / water, will help in identifying their coexistence on the icy mantles of cold dust grains. Experiment recording the IR spectra of methyl propionate as a function of temperature will be carried out as a separate spectroscopy study, which is beyond the scope of this manuscript, which will explain the nature of dimer formation, the phase transition and the present phase of methyl propionate at 85 K. Theoretical modelling strongly suggests the presence of methyl propionate molecules in the ice phase which will be a precursor for even larger and complex molecules.

## ACKNOWLEDGEMENTS

BS would like to acknowledge Prof N J Mason, The Open University, UK and R Mukherjee, PRL, Ahmedabad, India for their valuable discussions and comments in improving the quality of this manuscript. GG and MH would like to thank JST, CREST entitled "Creation of Innovative Functions of Intelligent Materials in the basis of the Element Strategy". AD and SKC want to thank ISRO

Respond (Grant No. ISRO/RES/2/372/11-12) and DST (Grant No. SB/S2HEP-021/2013) project for their financial supports. LM want to thank MOES for funding during this work. BS also thank INSPIRE grant (IFA- 11 CH -11).